\documentclass[letterpaper,conference,onecolumn,11pt]{ieeeconf}
\IEEEoverridecommandlockouts                              
\usepackage{srcltx}
\usepackage{color}
\usepackage{graphicx}
\usepackage{bm}
\usepackage[cmex10]{amsmath}
\usepackage{algorithmic}
\usepackage{algorithm}
\usepackage{amssymb}
\usepackage{multirow}
\usepackage{subfigure}
\newtheorem{remark}{Remark}

\usepackage{amsfonts}

\newtheorem{theorem}{Theorem}
\newtheorem{lemma}[theorem]{Lemma}
\newtheorem{definition}{Definition}
\newcounter{assumption}
\renewcommand{\theassumption}{A\arabic{assumption}}

\newenvironment{assumption}[1][]{\begin{trivlist}\item[] \refstepcounter{assumption}%
 \textbf{Assumption\ \theassumption\ #1} }{
 \ifvmode\smallskip\fi\end{trivlist}}

\newcommand{\HH}{\mathcal{H}}

\newcommand{\kfun}{\textsc{k}}

\newcommand{\beq}{\begin{equation}}
\newcommand{\eeq}{\end{equation}}
\newcommand{\beqa}{\begin{eqnarray}}
\newcommand{\eeqa}{\end{eqnarray}}
\newcommand{\beqan}{\begin{eqnarray*}}
\newcommand{\eeqan}{\end{eqnarray*}}
\newcommand{\ben}{\begin{eqnarray*}}
\newcommand{\een}{\end{eqnarray*}}

\newcommand{\smallnorm}[1]{\Vert#1\Vert}

\newcommand{\Real}{\mathbb R}

\newcommand{\EE}[1]{{\mathbb E}\left[#1\right]}

\newcommand{\ra}{\rightarrow}

\newcommand{\argmin}{\mathop{\textrm{argmin}}}

\newcommand{\eqdef}{\triangleq}

\newcommand{\Id}{\mathbf{I}}

\newcommand{\One}[1]{{\mathbb I}{\{#1\}}}

\newif\ifLongVersion
\LongVersiontrue

\begin{document}

\title{Learning-Based Modular Indirect Adaptive Control for a Class of Nonlinear Systems }
\author{Mouhacine Benosman, Amir-massoud Farahmand, Meng Xia
\thanks{ Mouhacine Benosman
(m{\_}benosman@ieee.org) is with the Multimedia group of
Mitsubishi Electric Research Laboratories (MERL), Cambridge, MA
02139, USA.  Amir-massoud Farahmand is with the Data Analytics
group of MERL. Meng Xia is with the with the Department of
Electrical Engineering, University of Notre Dame, Notre Dame, IN
46556, USA (part of this work has been completed during her
internship at MERL).} }

\maketitle

\begin{abstract}
We study in this paper the problem of adaptive trajectory tracking
control for a class of nonlinear systems with parametric
uncertainties. We propose to use {\it a modular approach}, where
we first design a robust nonlinear state feedback which renders
the closed loop input-to-state stable (ISS), where the input is
considered to be the estimation error of the uncertain parameters,
and the state is considered to be the closed-loop output tracking
error. Next, we augment this robust ISS controller with a
model-free learning algorithm to estimate the model uncertainties.
We implement this method with two different learning approaches.
The first one is a model-free multi-parametric extremum seeking
(MES) method and the second is a Bayesian optimization-based
method called Gaussian Process Upper Confidence Bound (GP-UCB).
The combination of the ISS feedback and the learning algorithms
gives a {\it learning-based modular indirect adaptive controller}.
We show the efficiency of this approach on a two-link robot
manipulator example.
\end{abstract}
\section{Introduction}
Classical adaptive methods can be classified into two main
approaches; `direct approaches', where the controller is updated
to adapt to the process, and `indirect approaches', where the
model is updated to better reflect the actual process. Many
adaptive methods have been proposed over the years for linear and
nonlinear systems, we could not possibly cite here all the design
and analysis results that have been reported, instead we refer the
reader to e.g., \cite{IS12,KKK95} and the references therein for
more details. Of particular interest to us is the indirect modular
approach to adaptive nonlinear control, e.g.,
\cite{KKK95,Wang06,benosmanatincacc13,benosmanatincecc13,benosmanatinccdc13,ben_ECC14,ben_ifac14,WL11,ben_ECC15,L09,HA011}.
In this approach, first the controller is designed by assuming
that all the parameters are known and then an identifier is used
to guarantee certain boundedness of the estimation error. The
identifier is independent of the designed controller and thus the
approach is called `modular'. For example, a modular approach has
been proposed in \cite{Wang06} for adaptive neural control of
pure-feedback nonlinear systems, where the input-to-state
stability (ISS) modularity of the controller-estimator is achieved
and the closed-loop stability is guaranteed by the small-gain
theorem (see also \cite{Sontag89,Jiang94}).\\In this work, we
build upon this type of modular adaptive design and provide a
framework which combines model-free learning methods and robust
model-based nonlinear control to propose a learning-based modular
indirect adaptive controller, where model-free learning algorithms
are used to estimate, in closed-loop, the uncertain parameters of
the model. The main difference with the existing model-based
indirect adaptive control methods, is the fact that we do not use
the model to design the uncertainty parameters estimation filters.
Indeed, model-based indirect adaptive controllers are based on
parameters estimators designed using the system's model, e.g., the
X-swapping methods presented in \cite{KKK95}, where gradient
descent filters obtained using the systems dynamics are designed
to estimate the uncertain parameters. We argue that because we do
not use the system's dynamics to design uncertainties estimation
filters we have less restrictions on the type of uncertainties
that we can estimate, e.g., uncertainties appearing nonlinearly
can be estimated with the proposed approach, see
\cite{benosmanatincecc13} for some earlier results on a
mechatronics application. We also show here that with the proposed
approach we can estimate at the same time a vector of linearly
dependent uncertainties, a case which cannot be straightforwardly
solved using model-based filters, e.g., refer to \cite{BA15} where
it is shown that the X-swapping model-based method fails to
estimate a vector of linearly dependent model coefficients. In
this work, we implement the proposed approach with two different
model-free learning algorithms: The first one is a dither-based
MES algorithm, and the second one is a Bayesian optimization-based
method called GP-UCB. The latter solves the
exploration-exploitation problem in the continuous armed bandit
problem, which is a non-associative reinforcement learning (RL)
setting. Indeed, MES is a model-free control approach with well
known convergence properties, since it has been analyzed in many
papers, e.g., \cite{Ariy03,Rote00,Ariy03,KW00,Teel01}. This makes
MES a good candidate for the model-free estimation part of our
modular adaptive controller, as already shown in some of our
preliminary results in \cite{ben_ECC14,ben_ifac14,ben_ECC15}.
However, one of the main limitations with dither-based MES is the
convergence to local minima. To improve this part of the
controller, we introduce here another model-free learning
algorithm in the estimation part of the adaptive controller.
Indeed, we propose in this paper to use a reinforcement learning
algorithm based on Bayesian optimization methods, known as GP-UCB,
e.g., \cite{SrinivasKrauseKakadeSeeger2010}, which contrary to the
MES algorithm is guaranteed to reach the global minima in a finite
search space.
\\One point worth mentioning at this stage is that comparatively
to`pure' model-free controllers, e.g., pure MES or model-free RL
algorithms, the proposed control here has a different goal.
Indeed, the available model-free controllers are meant for output
or state regulation, i.e., solving a static optimization problem.
In the contrary, here we propose to use model-free learning to
complement a model-based nonlinear control to estimate the unknown
parameters of the model, which means that the control goal, i.e.,
state or output trajectory tracking is handled by the model-based
controller. The learning algorithm is used to improve the tracking
performance of the model-based controller, and once the learning
algorithm has converged, one can carry on using the nonlinear
model-based feedback controller alone, i.e., without the need of
the learning algorithm. Furthermore, due to the fact that we are
merging together a model-based control with a model-free learning
algorithm, we believe that this type of controller can converge
faster to an optimal performance, comparatively to the pure
model-free controller, since by `partly' using a model-based
controller, we are taking advantage of the partial information
given by the physics of the system, whereas the pure model-free
algorithms assume no knowledge about the system, and thus start
the search for an optimal control signal from scratch.
\\ Similar ideas of merging model-based control and MES has been proposed
in
\cite{HA011,HA13,ben_patent12,benosmanatincacc13,benosmanatincecc13,benosmanatinccdc13,ben_ECC14,ben_ifac14,ben_ECC15}.
For instance in \cite{HA011,HA13}  extremum seeking is used to
complement a model-based controller, under linearity of the model
assumption in \cite{HA011} (in the direct adaptive control
setting, where the controllers gains are estimated), or in the
indirect adaptive control setting, under the assumption of linear
parametrization of the control in terms of the uncertainties in
\cite{HA13}. The modular design idea of using a model-based
controller with ISS guaranty, complemented with an MES-based
module can be found in
\cite{benosmanatincecc13,benosmanatinccdc13,ben_ECC14,ben_ifac14,ben_ECC15},
where the MES was used to estimate the model parameters and in
\cite{benosmanatincacc13,Benosman_book_chap15}, where feedback
 gains were tuned using MES algorithms. The work of this paper falls in this class of ISS-based modular
indirect adaptive controllers. The difference with other MES-based
adaptive controllers is that, due to the ISS modular design we can
use any model-free learning algorithm to estimate the model
uncertainties, not necessarily extremum seeking-based. To emphasis
this we show here the performance of the controller when using a
type of RL-based learning algorithm as well.
\\The rest of the paper is organized as follows. In Section
\ref{sec:pre}, we present some notations, and fundamental
definitions that will be needed in the sequel. In Section
\ref{sec:problem}, we formulate the problem. The nominal
controller design are presented in Section \ref{sec:norm}. In
Section \ref{sec:robust}, a robust controller is designed which
guarantees ISS from the estimation error input to the tracking
error state. In Section \ref{sec:ES}, the ISS controller is
complemented with an MES algorithm to estimate the model
parametric uncertainties. In section \ref{sec:GP-UCB}, we
introduce the RL GP-UCB algorithm as a model-free learning to
complement the ISS controller. Section \ref{sec:sim} is dedicated
to an application example and the paper conclusion is given in
Section \ref{sec:con}.

\section{Preliminaries}
\label{sec:pre} Throughout the paper, we use $\|\cdot\|$ to denote
the Euclidean norm; i.e., for a vector $x\in\mathbb{R}^n$, we have
$\|x\| \triangleq\|x\|_2 =\sqrt{x^Tx}$, where $x^T$ denotes the
transpose of the vector $x$.  We denote by $\textrm{Card}(S)$ the
size of a finite set $S$. The Frobenius norm of a matrix $A\in
\mathbb{R}^{m\times n}$, with elements $a_{ij}$, is defined as
$\|A\|_{F}\triangleq\sqrt{\sum_{i=1}^{n}\sum_{j=1}^{n}|a_{ij}|^{2}}$.
Given $x\in\mathbb{R}^m$, the signum function is defined as $
{\rm{sign} } (x)~\triangleq~[{\rm{sign} } (x_1),~{\rm{sign} }
(x_2),~\cdots,~{\rm{sign} } (x_m)]^T, $
where $sign(.)$ denotes the classical signum function. We use
$\dot{f}$ to denote the time derivative of $f$ and $f^{(r)}(t)$
for the $r$-th derivative of $f(t)$, i.e.
$f^{(r)}~\triangleq~\frac{d^r f}{dt}$. We denote
  by $\mathbb{C}^k$, functions that are $k$ times differentiable and by $\mathbb{C}^\infty$, a smooth function. A continuous
  function $\alpha : [0,a)\rightarrow [0,\infty)$ is said to belong to class $\mathcal{K}$ if it is strictly increasing
  and $\alpha(0) = 0$. It is said to belong to class $\mathcal{K}_{\infty}$ if $a = \infty$ and $\alpha(r)\rightarrow \infty$
  as $r\rightarrow \infty$ \cite{Khalil02}. A continuous function $\beta:[0,a)\times[0,\infty)\rightarrow [0,\infty)$ is said
  to belong to class $\mathcal{KL}$ if, for a fixed
$s$, the mapping $\beta(r,s)$ belongs to class $\mathcal{K}$ with respect to $r$ and, for each fixed $r$, the mapping $\beta(r,s)$
is decreasing with respect to $s$ and $\beta(r,s)\rightarrow 0$ as $s\rightarrow \infty$ \cite{Khalil02}.
\\Next, We introduce some
definitions that will be used in the sequel, e.g. \cite{Khalil02}:
Consider the system
\begin{align}
\label{eq:sys} \dot{x} ~=~f(t,x,u)
\end{align}
where $f: [0,\infty)\times
\mathbb{R}^n\times\mathbb{R}^m\rightarrow \mathbb{R}^n$ is
piecewise continuous in $t$ and locally Lipschitz in $x$ and $u$,
uniformly in $t$. The input $u(t)$ is piecewise continuous,
bounded function of $t$ for all $t\geq0$.

\begin{definition}[\cite{Khalil02,Mali05}]
\label{def:ISS} The system (\ref{eq:sys}) is said to be
\emph{input-to-sate stable} (ISS) if there exist a class
$\mathcal{KL}$ function $\beta$ and a class $\mathcal{K}$ function
$\gamma$ such that for any initial state $x(t_0)$ and any bounded
input $u(t)$, the solution $x(t)$ exists for all $t\geq t_0$ and
satisfies
\begin{align*}
\|x(t)\|\leq \beta(\|x(t_0)\|,t-t_0) +
\gamma(\sup_{t_0\leq\tau\leq t} \|u(\tau)\|).
\end{align*}
\end{definition}

\begin{theorem}[\cite{Khalil02,Mali05}]
\label{thm:iss} Let $V: [0,\infty)\times\mathbb{R}^n\rightarrow
\mathbb{R}$ be a continuously differentiable function such that
\begin{align}
\label{eq:iss}
\alpha_1(\|x\|) \leq& V(t,x)\leq \alpha_2(\|x\|)\notag\\
\frac{\partial V}{\partial t} + \frac{\partial V}{\partial
x}f(t,x,u)&\leq -W(x),\quad \forall \|x\|\geq\rho(\|u\|)>0
\end{align}
for all $(t,x,u)\in[0,\infty)\times\mathbb{R}^n\times
\mathbb{R}^m$, where $\alpha_1$, $\alpha_2$ are class
$\mathcal{K}_{\infty}$ functions, $\rho$ is a class $\mathcal{K}$
function, and $W(x)$ is a continuous positive definite function on
$\mathbb{R}^n$. Then, the system (\ref{eq:sys}) is input-to-state
stable (ISS).
\end{theorem}

\begin{remark}
Note that other equivalent definitions for ISS have been given in
\cite[pp. 1974-1975]{Mali05}. For instance, Theorem \ref{thm:iss}
holds if inequality (\ref{eq:iss}) is replaced by
\begin{align*}
\frac{\partial V}{\partial t} + \frac{\partial V}{\partial
x}f(t,x,u)&\leq -\mu(\|x\|) + \Omega(\|u\|)
\end{align*}
where $\mu\in\mathcal{K}_{\infty}\bigcap C^1$ and
$\Omega\in\mathcal{K}_{\infty}$.
\end{remark}

\section{Problem Formulation}
\label{sec:problem}
\subsection{Nonlinear system model}
We consider here affine uncertain nonlinear systems of the form
\begin{equation}
\begin{array}{l}
\displaystyle \dot{x} ~=~ f(x) + \Delta f(t,x) + g(x)u,\\
\displaystyle y~=~ h(x),
\end{array}
\label{eq:model}
\end{equation}
where $x\in\mathbb{R}^n$, $u\in \mathbb{R}^p$, $y\in\mathbb{R}^m$
($p~\geq~ m$), represent the state, the input and the controlled
output vectors, respectively. $\Delta f(t,x)$ is a vector field
representing additive model uncertainties. The vector fields $f$,
$\Delta f$, columns of $g$ and function $h$ satisfy the following
assumptions.

\begin{assumption}\label{ass1}
The function $f:\mathbb{R}^n \rightarrow \mathbb{R}^n$ and the
columns of $g: \mathbb{R}^n \rightarrow \mathbb{R}^p$ are
$\mathbb{C}^{\infty}$ vector fields on a bounded set $X$ of
$\mathbb{R}^n$ and $h : \mathbb{R}^n \rightarrow \mathbb{R}^m$ is
a $\mathbb{C}^{\infty}$ vector on $X$. The vector field $\Delta
f(x) $ is $\mathbb{C}^1$ on $X$.
\end{assumption}

\begin{assumption}\label{ass2}
System (\ref{eq:model}) has a well-defined (vector) relative
degree $\{r_1,~r_2,~\cdots,~r_m\}$ at each point $x^0\in X$, and
the system is linearizable, i.e., $\sum_{i=1}^m r_i=n$.
\end{assumption}


\begin{assumption}\label{ass3}
The desired output trajectories $y_{id}$ ($1\leq i \leq m$) are
smooth functions of time, relating desired initial points
$y_{id}(0)$ at $t=0$ to desired final points $y_{id}(t_f)$ at
$t=t_f$.
\end{assumption}

\subsection{Control objectives}
Our objective is to design a state feedback adaptive controller
such that the output tracking error is uniformly bounded, whereas
the tracking error upper-bound is function of the uncertain
parameters estimation error, which can be decreased by the
model-free learning. We stress here that the goal of learning
algorithm is not stabilization but rather performance
optimization, i.e., the learning improves the parameters
estimation error, which in turn improves the output tracking
error. To achieve this control objective, we proceed as follows:
First, we design a robust controller which can guarantee
input-to-state stability (ISS) of the tracking error dynamics
w.r.t the estimation errors input. Then, we combine this
controller with a model-free learning algorithm to iteratively
estimate the uncertain parameters, by optimizing online a desired
learning cost function.

\section{Adaptive Controller Design}
\label{sec:norm}
\subsection{Nominal Controller}
Let us first consider the system under nominal conditions, i.e.,
when $\Delta f(t,x) = 0$. In this case, it is well know, e.g.,
\cite{Khalil02}, that system (\ref{eq:model}) can be written as
\begin{align}
\label{eq:linear}
y^{(r)}(t) ~=~ b(\xi(t)) + A (\xi(t)) u(t),
\end{align}
where
\begin{equation}
\begin{array}{l}
\displaystyle y^{(r)}(t) ~=~ [y^{(r_1)}_{1}(t), ~y^{(r_2)}_{2}(t),~\cdots,~ y^{(r_m)}_{m}(t)]^T,\\
\displaystyle \xi(t) ~=~ [\xi^1(t),~\cdots,~\xi^m(t)]^T,\\
\displaystyle \xi ^i(t) ~=~[y_i(t),~\cdots,~y_i^{(r_i-1)}(t)], \quad 1\leq i\leq m
\end{array}
\label{eq:linear_c}
\end{equation}
The functions $b(\xi)$, $A(\xi)$ can be written as functions of
$f$, $g$ and $h$, and $A(\xi)$ is non-singular in $\tilde{X}$,
where $\tilde X$ is the image of the set of $X$ by the
diffeomorphism $x\rightarrow\xi$ between the states of system
(\ref{eq:model}) and the linearized model (\ref{eq:linear}). Now,
to deal with the uncertain model, we first need to introduce one
more assumption on system (\ref{eq:model}).

\begin{assumption}\label{lin_dist}
The additive uncertainties $\Delta f(t,x)$ in (\ref{eq:model})
appear as additive uncertainties in the input-output linearized
model (\ref{eq:linear})-(\ref{eq:linear_c}) as follows (see also
\cite{Beno09})
\begin{align}
\label{eq:model_un}
y^{(r)}(t) ~=~ b(\xi(t)) + A (\xi(t)) u(t) + \Delta b(t,\xi(t)),
\end{align}
where $\Delta b(t,\xi)$ is $\mathbb{C}^1$ w.r.t. the state vector
$\xi\in\tilde X$.
\end{assumption}

\begin{remark}
Assumption \ref{lin_dist}, can be ensured under the so-called
matching conditions (\cite{EO92}, p. 146).

\end{remark}

It is well known that the nominal model (\ref{eq:linear}) can be
easily transformed into a linear input-output mapping. Indeed, we
can first define a virtual input vector $v(t)$ as
\begin{align}
\label{eq:input}
v(t) ~=~ b(\xi(t)) + A (\xi(t)) u(t).
\end{align}
Combining (\ref{eq:linear}) and (\ref{eq:input}), we can obtain the following input-output mapping
\begin{align}
\label{eq:virtual}
y^{(r)}(t) ~=~v(t).
\end{align}
Based on the linear system (\ref{eq:virtual}), it is
straightforward to design a stabilizing controller for the nominal
system (\ref{eq:linear}) as
\begin{align}
\label{eq:norm}
u_{n} = A^{-1}(\xi) \left[v_{s}(t,\xi) - b(\xi)\right],
\end{align}
where $v_s$ is a $m\times1$ vector and the $i$-th ($1\leq i \leq m$) element $v_{si}$ is given by
\begin{align}
\label{eq:feedback}
v_{si} = y_{id}^{(r_i)} - K_{r_i}^{i}(y_i^{(r_i-1)}-y_{id}^{(r_i-1)})-\cdots-K_{1}^{i}(y_i-y_{id}).
\end{align}
If we denote the tracking error as $e_i(t) ~\triangleq~ y_i(t)-
y_{id}(t)$, we obtain  the following tracking error dynamics
\begin{align}
\label{eq:error}
e_i^{(r_i)}(t) +  K^i_{r_i}e^{(r_i-1)}(t) + \cdots + K^i_1e_i(t) = 0,
\end{align}
where $i \in \{1,~2,~\cdots,~m\}$. By properly selecting the gains
$K^i_j$ where $i \in \{1,~2,~\cdots,~m\}$ and
$j\in\{1,~2,~\cdots,~r_i\}$, we can obtain global asymptotic
stability of the tracking errors $e_i(t)$. To formalize this
condition, we add the following assumption.

\begin{assumption}\label{ass5}
There exists a non-empty set $\mathcal{A}$ where
$K^i_j\in\mathcal{A} $ such that the polynomials in
(\ref{eq:error}) are Hurwitz, where $i \in \{1,~2,~\cdots,~m\}$
and $j\in\{1,~2,~\cdots,~r_i\}$.
\end{assumption}

To this end, we define $z=[z^1,~z^2,~\cdots,~z^m]^T$, where $z^i=[e_i,~\dot{e_i},~\cdots,~e_i^{(r_i-1)}]$ and $i\in\{1,~2,\cdots,~m\}$. Then, from (\ref{eq:error}), we can obtain
\begin{align*}
\dot{z}~=~ \tilde{A} z,
\end{align*}
where $\tilde{A}\in \mathbb{R}^{n\times n}$ is a diagonal block matrix given by
\begin{align}
\label{eq:tildA}
\tilde{A}~=~{\rm{diag}}\{\tilde A_1,~\tilde A_2,~\cdots,~\tilde A_m\},
\end{align}
and $\tilde A_i$ ($1\leq i\leq m$) is a $r_i\times r_i$ matrix given by
\[ \tilde{A}_i~=~\left[ \begin{array}{cccccc}
0& 1&  \\
0& &1&\\
0&&&\ddots\\
\vdots&&&&1\\
-K^i_1\quad&-K^i_2\quad&\cdots\quad&\cdots\quad&-K^i_{r_i}
\end{array} \right].\]
As discussed above, the gains $K_j^i$ can be chosen such that the
matrix $\tilde{A}$ is Hurwitz. Thus, there exists a positive
definite matrix $P~>~ 0$ such that (see e.g. \cite{Khalil02})
\begin{align}
\label{eq:lyap}
\tilde A^T P + P \tilde A ~=~-I.
\end{align} 

In the next section, we build upon the nominal controller
(\ref{eq:norm}) to write a robust ISS controller.

\subsection{Lyapunov reconstruction-based ISS Controller} \label{sec:robust} We now consider the uncertain model
(\ref{eq:model}), i.e., when $\Delta f(t,x)\neq0$. The
corresponding exact linearized model is given by
(\ref{eq:model_un}) where $\Delta b (t,\xi(t))\neq0$. The global
asymptotic stability of the error dynamics (\ref{eq:error}) cannot
be guaranteed anymore due to the additive uncertainty $\Delta
b(t,\xi(t))$. We use  Lyapunov reconstruction techniques to design
a new controller so that the tracking error is guaranteed to be
bounded given that the estimate error of $\Delta b(t,\xi(t))$ is
bounded. The new controller for the uncertain model
(\ref{eq:model_un}) is defined as
\begin{align}
\label{eq:full}
u_{f}~=~u_{n} + u_{r},
\end{align}
where the nominal controller $u_{n}$ is given by (\ref{eq:norm})
and the robust controller $u_r$ will be given later. By using the
controller (\ref{eq:full}), and (\ref{eq:model_un}) we obtain
\begin{align}
\label{eq:io_new}
y^{(r)}(t) &~=~ b(\xi(t)) + A (\xi(t)) u_{f} + \Delta b(t,\xi(t)),\notag\\
&~=~ b(\xi(t)) + A (\xi(t)) u_{n} + A (\xi(t)) u_{r}+ \Delta b(t,\xi(t)),\notag\\
& ~=~v_{s}(t,\xi)+ A (\xi(t)) u_{r}+ \Delta b(t,\xi(t)),
\end{align}
where (\ref{eq:io_new}) holds from (\ref{eq:norm}). Which leads to
the following error dynamics
\begin{align}
\label{eq:dyn_z}
\dot{z}~=~ \tilde{A} z + \tilde{B} \delta,
\end{align}
where $\tilde{A}$ is defined in (\ref{eq:tildA}), $\delta$ is a $m\times 1$ vector given by
\begin{align}
\label{eq:delta}
\delta~=~A (\xi(t)) u_{r}+ \Delta b(t,\xi(t)),
\end{align}

and the matrix $\tilde B\in\mathbb{R}^{n\times m}$ is given by
\begin{align}
\label{eq:tildB}
\tilde{B}~=~\left[ \begin{array}{c}
\tilde{B}_1 \\
\tilde{B}_2\\
\vdots\\
\tilde{B}_m
\end{array} \right],
\end{align}
where each $\tilde B_i$ ($1\leq i\leq m$) is given by a $r_i\times
m$ matrix such that
\[ \tilde{B}_i(l,q) ~=~ \left\{ \begin{array}{lll}
         1 & \quad\mbox{for $l = r_i$, $q = i$}\\
         0  & \quad\mbox{otherwise.}\end{array} \right. \]

If we choose $V(z)=z^TPz$ as a Lyapunov function for the dynamics
(\ref{eq:dyn_z}), where $P$ is the solution of the Lyapunov
equation (\ref{eq:lyap}), we obtain
\begin{align}
\label{eq:general}
\dot{V}(t)&~=~\frac{\partial V}{\partial z}\dot{z},\notag\\
&~=~z^T(\tilde A^T P + P \tilde A)z + 2 z^T P \tilde B \delta,\notag\\
&~=~-\|z\|^2 + 2 z^T P \tilde B \delta,
\end{align}

where $\delta$ given by (\ref{eq:delta}) depends on the robust controller $u_r$.

Next, we design the controller $u_{r}$ based on the form of the
uncertainties $\Delta b(t,\xi(t))$.  More specifically, we
consider here the case when $\Delta b (t,\xi(t))$ is of the
following form
\begin{align}
\label{eq:affine} \Delta b (t,\xi(t))~=~E\;Q(\xi,t),
\end{align}
where $E\in\mathbb{R}^{m\times m}$ is a matrix of unknown constant
parameters, and
$Q(\xi,t):\;\mathbb{R}^{n}\times\mathbb{R}\rightarrow\mathbb{R}^{m}$
is a known bounded function of sates and time variables. For
notational convenience, we denote by $\hat{E}(t)$ the estimate of
$E$, and by $e_{E}=E-\hat{E}$, the estimate error. We define the
unknown parameter vector
$\Delta=[E(1,1),...,E(m,m)]^{T}\in\mathbb{R}^{m^{2}}$, i.e.,
concatenation of all elements of $E$, its estimate is denoted by
$\widehat{\Delta}(t)=[\hat{E}(1,1),...,\hat{E}(m,m)]^{T}$, and the
estimation error vector is given by $e_\Delta(t) =
\Delta-\widehat{\Delta}(t)$.\\

Next, we propose the following robust controller
\begin{align}
\label{eq:case3} u_{r}~=~&-A^{-1}(\xi)[\tilde B^TPz\|Q(\xi,t)\|^2
+\widehat{E}(t)Q(\xi,t)].
\end{align}
The closed-loop error dynamics can be written as
\begin{align}
\label{eq:cl_case2} \dot{z}~=~\tilde{f}(t,z,e_\Delta),
\end{align}
where $e_\Delta(t)$ is considered to be an input to the system
(\ref{eq:cl_case2}).

\begin{theorem}
\label{thm:case3} Consider the system (\ref{eq:model}), under
Assumptions \ref{ass1}-\ref{ass5}, where $\Delta b(t,\xi(t))$
satisfies (\ref{eq:affine}). If we apply to (\ref{eq:model}) the
feedback controller (\ref{eq:full}), where $u_n$ is given by
(\ref{eq:norm}) and $u_r$ is given by (\ref{eq:case3}). Then, the
closed-loop system (\ref{eq:cl_case2}) is ISS from the estimation
errors input $e_\Delta(t)\in\mathbb{R}^{m^{2}}$ to the tracking
errors state $z(t)\in\mathbb{R}^n$.
\end{theorem}
{\it Proof:} By substitution (\ref{eq:case3}) into
(\ref{eq:delta}), we obtain
\begin{align*}
\delta=&~-\tilde B^TPz\|Q(\xi,t)\|^2 -\widehat{E}(t)\; Q(\xi,t) + \Delta b(t,\xi(t))\\
=&~-\tilde B^TPz\|Q(\xi,t)\|^2 -\widehat{E}(t)\; Q(\xi,t) + E\;
Q(\xi,t),
\end{align*}
If we consider $V(z)=z^TPz$ as a Lyapunov function for the error
dynamics (\ref{eq:dyn_z}). Then, from (\ref{eq:general}), we
obtain
\begin{align*}
\dot{V}\leq & -\|z\|^2 + 2 z^T P \tilde B E\; Q(\xi,t)- 2 z^T P \tilde B\widehat{E}(t)\; Q(\xi,t)\\
&-2\|z^T P \tilde B\|^2\|Q(\xi,t)\|^2,
\end{align*}
which leads to
\begin{align*}
\dot{V} \leq~& -\|z\|^2+2 z^T P \tilde B e_{E} Q(\xi,t)-2\|z^T P
\tilde B\|^2\|Q(\xi,t)\|^2.
\end{align*}
Since $z^T P \tilde B e_{E} Q(\xi)\leq \|z^T P \tilde B e_{E}
Q(\xi)\| \leq \|z^T P \tilde B\|\| e_{E}\|_{F}\| Q(\xi)\|=\|z^T P
\tilde B\|\| e_{\Delta}\|\| Q(\xi)\|$, we obtain
\begin{align*}
\dot{V} \leq~& -\|z\|^2+2 \|z^T P \tilde B\|\| e_\Delta\| \|Q(\xi,t)\|-2\|z^T P \tilde B\|^2\|Q(\xi,t)\|^2\\
~\leq~& -\|z\|^2 -2(\|z^T P \tilde B\|\|Q(\xi,t)\|-\frac{1}{2}\|e_\Delta\|)^2 + \frac{1}{2}\|e_\Delta\|^2\notag\\
~\leq~&-\|z\|^2+\frac{1}{2}\|e_\Delta\|^2.
\end{align*}
Thus, we have the following relation
\begin{align*}
\dot{V}\leq-\frac{1}{2}\|z\|^2,\quad \forall
\|z\|\geq\|e_\Delta\|>0,
\end{align*}
Then from the Lyapunov direct theorem in  \cite{Khalil02,Mali05},
we obtain that system (\ref{eq:cl_case2}) is ISS from input
$e_\Delta$ to state $z$.
\\
\subsection{MES-based parametric uncertainties estimation} \label{sec:ES}
Let us define now the following cost function
\begin{align}
\label{eq:cost_gen} J(\widehat{\Delta}) ~=~
F(z(\widehat{\Delta})),
\end{align}
where $F : \mathbb{R}^{n} \rightarrow \mathbb{R}$, $F({\bf 0}) =
0$, $F(z)>0$ for $z\in\mathbb{R}^{n}-\{\bf 0\}$. We need the
following assumptions on $J$.
\begin{assumption}
\label{asp:cost}
The cost function $J$ has a local minimum at $\widehat{\Delta}^* ~=~\Delta$.
\end{assumption}

\begin{assumption}
\label{asp:local} The initial error $e_\Delta(t_0)$ is
sufficiently small, i.e., the original parameter estimate vector
$\widehat{\Delta}$ are close enough to the actual parameter vector
$\Delta$.
\end{assumption}

\begin{assumption}
\label{asp:Lips} The cost function $J$ is analytic and its
variation with respect to the uncertain parameters is bounded in
the neighborhood of $\widehat{\Delta}^*$, i.e., $\|\frac{\partial
J}{\partial \widehat{\Delta}}(\tilde \Delta)\|\leq \xi_2$,
$\xi_2>0$, $\tilde{\Delta}\in\mathcal{V}(\widehat{\Delta}^*)$,
where $\mathcal{V}(\widehat{\Delta}^*)$ denotes a compact
neighborhood of $\widehat{\Delta}^*$.
\end{assumption}



We can now present the following result.
\begin{lemma}
\label{lem:case1} Consider the system (\ref{eq:model}), under
Assumptions \ref{ass1}-\ref{asp:Lips}, where the uncertainty is
given by (\ref{eq:affine}). If we apply to (\ref{eq:model}) the
feedback controller (\ref{eq:full}), where $u_n$ is given by
(\ref{eq:norm}), $u_r$ is given by (\ref{eq:case3}), the cost
function is given by (\ref{eq:cost_gen}), and
$\widehat{\Delta}(t)$ are estimated through the ES algorithm
\begin{align}
\label{eq:mes}
\dot{\tilde{x}}_{i}~&=~a_{i} \sin(\omega_{i} t+ \frac{\pi}{2})J(\widehat{\Delta}),\;a_{i}>0,\notag\\
\widehat{\Delta}_{i}(t)~&=~\tilde{x}_{i}+a_{i}\sin(\omega_{i} t -
\frac{\pi}{2}), \quad i\in\{1,2,\dots,m^{2}\}
\end{align}
with $\omega_{i}\neq \omega_{j}$, $\omega_{i} +
\omega_{j}\neq\omega_{k}$, $i,j,k\in\{1,2,\dots,m^{2}\}$, and
$\omega_{i}>\omega^*$, $\forall ~i\in\{1,2,\dots,m^{2}\}$, with
$\omega^*$ large enough. Then, the norm of the error vector $z(t)$
admits the following bound
\begin{align*}
\|z(t)\|&\leq\beta(\|z(0)\|,t) +
\gamma(\tilde{\beta}(\|e_\Delta(0)\|,t)+\|e_\Delta\|_{\max}),
\end{align*}
where $\|e_\Delta\|_{\max} = \frac{\xi_1}{\omega_0} +
\sqrt{\sum_{i= 1}^{m^{2}} a_{i}^2}$, $\xi_1>0$, $\omega_0 =
\max_{i\in\{1,2,\dots,m^{2}\}} \omega_{i}$,
$\beta\in\mathcal{KL}$, $\tilde{\beta}\in\mathcal{KL}$ and
$\gamma\in\mathcal{K}$.
\end{lemma}

{\it Proof:} Based on Theorem \ref{thm:case3}, we know that the
tracking error dynamics (\ref{eq:cl_case2}) is ISS from the input
$e_\Delta(t)$ to the state $z(t)$. Thus, by Definition
\ref{def:ISS}, there exist a class $\mathcal{KL}$ function $\beta$
and a class $\mathcal{K}$ function $\gamma$ such that for any
initial state $z(0)$, any bounded input $e_\Delta(t)$ and any
$t\geq0$,
\begin{align}
\label{eq:bound} \|z(t)\|\leq \beta(\|z(0)\|,t) +
\gamma(\sup_{0\leq\tau\leq t} \|e_\Delta(\tau)\|).
\end{align}
Now, we need to evaluate the bound on the estimation vector
$\widehat{\Delta}(t)$, to do so we use the results presented in
\cite{Rote00}. First, based on Assumption \ref{asp:Lips}, the cost
function is locally Lipschitz, i.e. there exists $\eta_1>0$ such
that $|J(\Delta_1)-J(\Delta_2)|\leq\eta_1\|\Delta_1 -\Delta_2\|$,
for all $\Delta_1, \Delta_2 \in\mathcal{V}(\widehat{\Delta}^*)$.
Furthermore, since $J$ is analytic, it can be approximated locally
in $\mathcal{V}(\widehat{\Delta}^*)$ by a quadratic function, e.g.
Taylor series up to the second order. Based on this and on
Assumptions \ref{asp:cost} and \ref{asp:local}, we can obtain the
following bound (\cite[p. 436-437]{Rote00},\cite{BA14})
\begin{align*}
\|e_\Delta(t)\| - \|d(t)\| \leq \|e_\Delta(t)-
d(t)\|\leq\tilde{\beta}(\|e_\Delta(0),t\|) +
\frac{\xi_1}{\omega_0},
\end{align*}
where $\tilde{\beta}\in\mathcal{KL}$, $\xi_1>0$, $t\geq0$,
$\omega_0 = \max_{i\in\{1,2,\dots,m^{2}\}} \omega_{i}$, and $d(t)
= [a_{1}\sin(\omega_{1} t +
\frac{\pi}{2}),\dots,a_{m^{2}}\sin(\omega_{m^{2}}
t+\frac{\pi}{2})]^T$. We can further obtain that
\begin{align*}
\|e_\Delta(t)\|&\leq~ \tilde{\beta}(\|e_\Delta(0),t\|) + \frac{\xi_1}{\omega_0} + \|d(t)\|\\
&\leq~\tilde{\beta}(\|e_\Delta(0),t\|) + \frac{\xi_1}{\omega_0} +
\sqrt{\sum_{i= 1}^{m^{2}} a_{i}^2}.
\end{align*}
Together with (\ref{eq:bound}) yields the desired result.\\

\begin{remark}
The adaptive controller of Lemma \ref{lem:case1} uses the ES
algorithm (\ref{eq:mes}) to estimate the model parametric
uncertainties. One might ask the question: where is the famous
persistence of excitation (PE) condition here ? The answer can be
found in the examination of equation (\ref{eq:mes}). Indeed, the
ES algorithm uses as `input' the sinusoidal signals $a_{i}
\sin(\omega_{i} t+ \frac{\pi}{2})$ which clearly satisfy the PE
condition. The main difference with classical adaptive control
result, is that these excitation signals are not entering the
system dynamics directly, but instead are applied as inputs to the
ES algorithm, reflected on the ES estimations outputs and thus
transmitted to the system through the feedback loop.
\end{remark}
As we mentioned earlier, the dither-based MES has a problem of
local minima, to improve this point we propose in the next section
to use GP-UCB as the model-free learning algorithm for model
uncertainties estimation.
\subsection{GP-UCB based parametric uncertainties estimation} \label{sec:GP-UCB}

\newcommand{\Deltahat}{\widehat{\Delta}}

In this section we propose to use Gaussian Process Upper
Confidence Bound (GP-UCB) algorithm to find the uncertain
parameter $\Delta$
vector~\cite{SrinivasKrauseKakadeSeeger2010,SrinivasKrauseKakadeSeeger2012}.
GP-UCB is a Bayesian optimization algorithm for stochastic
optimization, i.e., the task of finding the global optimum of an
unknown function when the evaluations are potentially contaminated
with noise. The underlying working assumption for Bayesian
optimization algorithms, including GP-UCB, is that the function
evaluation is costly, so we would like to minimize the number of
evaluations while having as accurate estimate of the minimizer (or
maximizer) as possible~\cite{BrochuCoradeFreitas2010}. For GP-UCB,
this goal is guaranteed by having an upper bound on the regret of
the algorithm -- to be defined precisely later.

One difficulty of stochastic optimization is that since we only
observe noisy samples from the function, we cannot really be sure
about the exact value of a function at any given point. One may
try to query a single point many times in order to have an
accurate estimate of the function. This, however, may lead to
excessive number of samples, and can be wasteful way of assigning
samples when the true value of the function at that point is
actually far from optimal. The Upper Confidence Bounds (UCB)
family of algorithms provide a principled approach to guide the
search~\cite{AuerCesaBianchiFischer2002}. These algorithms, which
are not necessarily formulated in a Bayesian framework,
automatically balance the exploration (i.e., finding regions of
the parameter space that \emph{might} be promising) and the
exploration (i.e., focusing on the regions that are known to be
the best based on the \emph{current} available knowledge) using
the principle of optimism in the face of uncertainty. These
algorithms often come with strong theoretical guarantee about
their performance. For more information about the UCB class of
algorithms, refer
to~\cite{BubeckMunosetal2011,BubeckCesaBianchi2012,MunosBanditsSurvery2014}.
GP-UCB is a particular UCB algorithms that is suitable to deal
with continuous domains. It uses a Gaussian Process (GP) to
maintain the mean and confidence information about the unknown
function.

We briefly discuss GP-UCB in our context following the discussion
of the original
papers~\cite{SrinivasKrauseKakadeSeeger2010,SrinivasKrauseKakadeSeeger2012}.
Consider the cost function $J: D \ra \Real$ to be minimized. This
function depends on the dynamics of the closed-loop system, which
itself depends on the parameters $\Deltahat$ used in the
controller design. So we may consider it as an unknown function of
$\widehat{\Delta}$.

For the moment, let us assume that $J$ is a function sampled from
a Gaussian Process (GP)~\cite{RasmussenWilliamsGP06}. Recall that
a GP is a stochastic process indexed by the set $D$ that has the
property that for any finite subset of the evaluation points, that
is $\{ \widehat{\Delta}_1, \widehat{\Delta}_2, \dotsc,
\widehat{\Delta}_t \} \subset D$, the joint distribution of
$\left( J(\widehat{\Delta}_i) \right)_{i=1}^t$ is a multivariate
Gaussian distribution. GP is defined by a mean function
$\mu(\widehat{\Delta}) = \EE{J(\widehat{\Delta})}$ and its
covariance function (or kernel)
$\kfun(\widehat{\Delta},\widehat{\Delta}') =
\textrm{Cov}(J(\widehat{\Delta}),J(\widehat{\Delta}')) = \EE{
\left( J(\widehat{\Delta}) - \mu(\widehat{\Delta}) \right)
\left(J(\widehat{\Delta}') - \mu(\widehat{\Delta}') \right) }$.
The kernel $\kfun$ of a GP determines the behavior of a typical
function sampled from the GP. For instance, if we choose
$\kfun(\widehat{\Delta},\widehat{\Delta}') = \exp\left( -
\frac{\smallnorm{\widehat{\Delta} - \widehat{\Delta}'}^2}{2 l^2}
\right)$, the squared exponential kernel with length scale $l >
0$, it implies that the the GP is mean square differentiable of
all orders.
We write $J \sim \text{GP}(\mu, \kfun)$.

Let us first briefly describe how we can find the posterior
distribution of a $\text{GP}(0, \kfun)$; a GP with zero prior
mean. Suppose that for $\underline{\Deltahat}_{t-1} \eqdef \{
\widehat{\Delta}_1, \widehat{\Delta}_2, \dotsc,
\widehat{\Delta}_{t-1} \} \subset D$, we have observed the noisy
evaluation $y_i = J(\widehat{\Delta}_i) + \eta_i$ with $\eta_i
\sim N(0,\sigma^2)$ being i.i.d. Gaussian noise. We can find the
posterior mean and variance for a new point $\Deltahat^* \in D$ as
follows: Denote the vector of observed values by $\mathbf{y}_{t-1}
= [y_1, \dotsc, y_{t-1}]^\top \in \Real^{t-1}$, and define the
Grammian matrix $K \in \Real^{t - 1 \times t - 1}$ with $[K]_{i,j}
= \kfun(\Deltahat_i,\Deltahat_j)$, and the vector $\kfun_* = [
\kfun(\underline{\Deltahat}_1,\Deltahat^*), \dotsc,
\kfun(\underline{\Deltahat}_{t-1},\Deltahat^*)]$. The expected
mean $\mu_{t}(\Deltahat^*)$ and the variance
$\sigma_t(\Deltahat^*)$ of the posterior of the GP evaluated at
$\Deltahat^*$ are (cf. Section 2.2
of~\cite{RasmussenWilliamsGP06})
\begin{align*}
    & \mu_{t}(\Deltahat^*) =
    \kfun_*  \left[ K + \sigma^2 \Id \right]^{-1} \mathbf{y}_{t-1},
    \\
    &
    \sigma_t^2({\Deltahat^*}) =
    \kfun(\Deltahat^*,\Deltahat^*) -
    \kfun_*^\top \left[ K + \sigma^2 \Id \right]^{-1} \kfun_*.
\end{align*}

At round $t$, the GP-UCB algorithm selects the next query point
$\Deltahat_t$ by solving the following optimization
problem\footnote{UCB algorithms are often formulated as
maximization problems, so the ``upper'' confidence bound is
calculated. Here we actually compute the ``lower'' confidence
bound, but to keep the naming convection, we still GP-UCB instead
of GP-LCB.}:

\begin{align}
\label{eq:IndirectAdaptiveControl-GPUCB-Optimization}
    \Deltahat_t \leftarrow \argmin_{\Deltahat \in D} \mu_{t-1}(\Deltahat) - \beta_t^{1/2} \sigma_{t-1}(\Deltahat).
\end{align}
Where $\beta_t$ depends on the choice of kernel among other
parameters of the problem.

The optimization
problem~\eqref{eq:IndirectAdaptiveControl-GPUCB-Optimization} is
often nonlinear and non-convex. Nonetheless solving it only
requires querying the GP, which in general is much faster than
querying the original dynamical system. This is important when the
dynamical system is a physical system and we would like to
minimize the number of interactions with it before finding a
$\Deltahat$ with small $J(\Deltahat)$. One practically easy way to
approximately
solve~\eqref{eq:IndirectAdaptiveControl-GPUCB-Optimization} is to
restrict the search to a finite subset $D'$ of $D$. The finite
subset can be a uniform grid structure over $D$, or it might
consist of randomly selected members of $D$. 

The theoretical guarantee for GP-UCB is in the form of regret
upper bound. Let us define $\Delta^* \leftarrow \argmin_{\Delta
\in D} J(\Delta)$, the global minimizer of the objective function.
The regret at time $t$ is defined by $r_t = J(\Deltahat_t) -
J(\Delta^*)$. This is a measure of sub-optimality of the choice of
$\Deltahat_t$ according the cost function $J$. The cumulative
regret at time $T$ is defined as $R_T = \sum_{t=1}^T r_t$. Ideally
we would like $\lim_{t \ra \infty} \frac{R_T}{T} = 0$.

The behavior of the cumulative regret $R_T$ depends on the set $D$
and the choice of kernel. If we fix the confidence parameter
$\delta
> 0$, for the squared exponential kernel, the asymptotic behavior of
$R_T$ is
\[
O \left( \sqrt{T [\log^{d+1}(T) + \log(1/\delta) ] } \right),
\]
with probability at least $1 - \delta$ (cf. Theorem~3
of\ifLongVersion~\cite{SrinivasKrauseKakadeSeeger2010,SrinivasKrauseKakadeSeeger2012}\else~\cite{SrinivasKrauseKakadeSeeger2010}\fi).
This result does not even require the function $J$ to be a GP. It
only requires the function to have a finite norm in the
reproducing kernel Hilbert space (RKHS) $\HH_\kfun$ defined by the
kernel $\kfun$.


%
%
%
%
%
%
%

%

\begin{remark}\label{remark1}
One main difference with some of the existing model-based adaptive
controllers, is the fact that the learning-based estimation
algorithm used here does not depend on the model of the system,
i.e., the only information needed to compute the learning cost
function (\ref{eq:cost_gen}) is the desired trajectory and the
measured output of the system (please refer to Section
\ref{sec:sim} for an example). This makes the learning-based
adaptive controllers suitable for the general case of nonlinear
parametric uncertainties. For example in \cite{ben_ECC13}, a
similar preliminary algorithm has been tested in the case of
nonlinear models of electromagnetic actuators with a nonlinear
parametric uncertainty. Another point worth mentioning here, is
the fact that with the available modular model-based adaptive
controllers, like the X-swapping modular algorithms, e.g.,
\cite{Krs95}, it is not possible in some cases to estimate
multiple uncertainties simultaneously. For instance, it is shown
in \cite{BA15} that the X-swapping adaptive control cannot
estimate multiple uncertainties in the case of electromagnetic
actuators, due to the linear dependency of the uncertain
parameters, i.e., when we consider three parametric uncertainties
affecting the same output acceleration, in which case the
model-based estimation filters cannot distinguish between the
uncertainties from this acceleration. However, when dealing with
the same example, the MES-based modular indirect adaptive control
approach was successful in estimating multiple uncertainties at
the same time \cite{BA14}. A similarly challenging case is
considered in the example presented in the next section.
\end{remark}

\section{Two-link Manipulator Example}
\label{sec:sim} We consider here a two-link robot manipulator,
with the following dynamics (see e.g. \cite{Spong92})
\begin{align}
\label{eq:robot}
H(q)\ddot{q} + C(q,\dot{q})\dot{q}+ G(q)~=~\tau,
\end{align}
where $q \triangleq [q_1,q_2]^T$ denotes the two joint angles and
$\tau \triangleq [\tau_1,\tau_2]^T$ denotes the two joint torques.
The matrix $H\in\mathbb{R}^{4\times 4}$ is assumed to be
non-singular and its elements are given by
\begin{equation}
\label{eq:robot_c}
\begin{array}{l}
\displaystyle H_{11} ~=~m_1\ell_{c_1}^2+I_1+m_2[\ell_1^2+\ell_{c_2}^2+ 2\ell_1\ell_{c_2}\cos(q_2)]+I_2, \\
\displaystyle H_{12} ~=~ m_2\ell_1\ell_{c_2}\cos(q_2)+m_2\ell_{c_2}^2+I_2,\\
\displaystyle H_{21} ~=~H_{12},\\
\displaystyle H_{22} ~=~ m_2\ell_{c_2}^2+I_2.
\end{array}
\end{equation}
The matrix $C(q,\dot{q})$ is given by
\[ C(q,\dot{q})~\triangleq~\left[ \begin{array}{ccc}
-h\dot{q_2} \quad& -h\dot{q_1}-h\dot{q_2} \\
h\dot{q_1}\quad&0\end{array} \right],\]
where $h~=~m_2\ell_1\ell_{c_2}\sin(q_2)$. The vector $G ~=~ [G_1,G_2]^T$ is given by
\begin{equation}
\begin{array}{l}
\displaystyle G_1 =~m_1\ell_{c_1}g\cos(q_1)+m_2 g[\ell_2\cos(q_1+q_2)+\ell_1\cos(q_1)],\\
\displaystyle G_2 =~ m_2\ell_{c_2}g\cos(q_1+q_2),
\end{array}
\label{eq:robot_g}
\end{equation}
where, $\ell_{1},\;\ell_{2}$ are the lengths of the first and
second link, respectively, $\ell_{c_1},\;\ell_{c_2}$ are the
distances between the rotation center and the center of mass of
the first and second link respectively. $m_{1},\;m_{2}$ are the
masses of the first and second link, respectively, $I_{1}$ is the
moment of inertia of the first link and $I_{2}$ the moment of
inertia of the second link, respectively, and $g$ denotes the
earth gravitational constant.\\ In our simulations, we assume that
the parameters take the following values: $I_2=\frac{5.5}{12}\;
kg\cdot m^2$, $m_1=10.5\;kg$, $m_2=5.5\;kg$, $\ell_1=1.1\;m$,
$\ell_2=1.1\;m$, $\ell_{c_1}=0.5\;m$, $\ell_{c_2}=0.5\;m$,
$I_1=\frac{11}{12}\;kg\cdot m^2$, $g=9.8\;m/s^2$. The system
dynamics (\ref{eq:robot}) can be rewritten as
\begin{align}
\label{eq:norm_robot} \ddot{q}~=~H^{-1}(q) \tau -  H^{-1}(q)
\left[C(q,\dot{q})\dot{q}+ G(q)\right].
\end{align}
Thus, the nominal controller is given by
\begin{align}
\label{tau_nominal_robot}
\tau_{n}~=&~\left[C(q,\dot{q})\dot{q}+ G(q)\right] \notag\\
&+~ H(q)\left[\ddot{q_{d}} -
K_{d}(\dot{q}-\dot{q_{d}})-K_{p}(q-q_{d})\right],
\end{align}
where $q_d =[q_{1d},q_{2d}]^T$, denotes the desired trajectory and
the diagonal gain matrices $K_{p}>0$, $K_{d}>0$, are chosen such
that the linear error dynamics (as in (\ref{eq:error})) are
asymptotically stable. We choose as output references the $5th$
order polynomials
$q_{1ref}(t)=q_{2ref}(t)=\sum_{i=0}^{5}a_{i}(t/t_{f})^{i}$, where
the $a_{i}$'s have been computed to satisfy the boundary
constraints
$q_{iref}(0)=0,q_{iref}(t_{f})=q_{f},\dot{q}_{iref}(0)=\dot{q}_{iref}(t_{f})=0,\ddot{q}_{iref}(0)=\ddot{q}_{iref}(t_{f})=0$,
$i=1,2$, with $t_{f}=2\;sec$, $q_{f}=1.5\;rad$. In these tests, we
assume that the nonlinear model (\ref{eq:robot}) is uncertain. In
particular, we assume that there exist additive uncertainties in
the model (\ref{eq:norm_robot}),  i.e.,
\begin{align}
\label{eq:uncertain_robot} \ddot{q}=~H^{-1}(q) \tau -  H^{-1}(q)
\left[C(q,\dot{q})\dot{q}+ G(q)\right]- E\; G(q).
\end{align}
Where, $E$ is a matrix of constant uncertain parameters. Following
(\ref{eq:case3}), the robust-part of the control writes as
\begin{equation}\label{tau_robust_robot}
\tau_{r}=-H(\tilde{B}^{T}Pz\|G\|^{2}-\hat{E}\; G(q)),
\end{equation}
where $$\tilde{B}^{T}=\left[
\begin{array}{cccc}
0&1&0&0\\0&0&0&1
\end{array}\right],
$$
$P$ is solution of the Lyapunov equation (\ref{eq:lyap}), with
$$
\tilde{A}=
\left[
\begin{array}{cccc}
0&1&0&0\\-K^{1}_{p}&-K^{1}_{d}&0&0\\0&0&0&1\\0&0&-K^{2}_{p}&-K^{2}_{d}
\end{array}
\right],
$$
$z=[q_{1}-{q}_{1d},\dot{q}_{1}-\dot{q}_{1d},q_{2}-{q}_{2d},\dot{q}_{2}-\dot{q}_{2d}]^{T}$,
and $\hat{E}$ is the matrix of the parameters' estimates.
Eventually, the final feedback controller writes as
\begin{equation}\label{full_controller_example}
\tau=\tau_{n}+\tau_{r}.
\end{equation}
We consider here the challenging case where the uncertain
parameters are linearly dependent. In this case the uncertainties'
`effect' is not observable from the measured output (see Remark
\ref{remark1}). Indeed, in the case where the uncertainties enter
the model in a linearly dependent function, e.g. when the matric
$\Delta$ has only one non-zero line, some of the classical
available modular model-based adaptive controllers, like for
instance X-swapping controllers, cannot be used to estimate all
the uncertain parameters simultaneously. For example, it has been
shown in \cite{BA15}, that the model-based gradient descent
filters failed to estimate simultaneously multiple parameters in
the case of the electromagnetic actuators example. For instance,
in comparison with the ES-based indirect adaptive controller of
\cite{HA13}, the modular approach does not rely on the parameters
mutual exhaustive assumption, i.e., each element of the control
vector needs to be linearly dependent on at least one element of
the uncertainties vector. More specifically, we consider here the
following case: $\Delta(1,1)=0.3,\;\Delta(1,2)=0.6$, and
$\Delta(2,i)=0,\;i=1,2.$ In this case, the uncertainties' effect
on the acceleration $\ddot{q}_{1}$ cannot be differentiated, and
thus the application of the model-based X-swapping method to
estimate the actual values of both uncertainties at the same time
is challenging. Similarly, the method of \cite{HA13}, cannot be
readily applied because the second control $\tau_{2}$ is not
linearly depend on the uncertainties, which only affects
$\tau_{1}$. However, we show next that, by using the modular
ISS-based controller, we manage to estimate the actual values of
the uncertainties simultaneously and improve the tracking
performance.
\subsection{MES-based uncertainties estimation}
\label{sec:constant} 
\begin{figure}\center
 \begin{minipage}{0.5\linewidth}
   \center\subfigure[Cost function over the learning iterations (MES)]{
\includegraphics[width=1\linewidth]{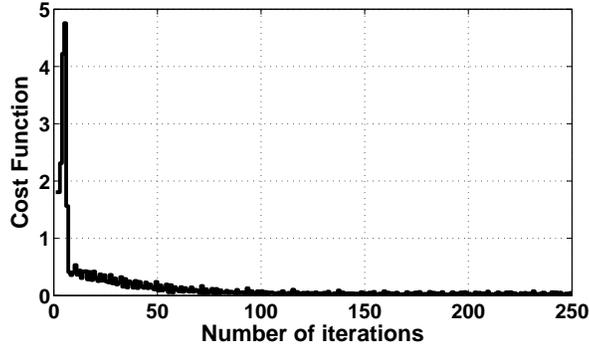}
\label{fig:cost_case2}}
  \end{minipage}
  \hfill
  \begin{minipage}{0.5\linewidth}
   \center\subfigure[Estimate of $ \Delta_{1}$ over the learning iterations (MES)]{
\includegraphics[width=1\linewidth]{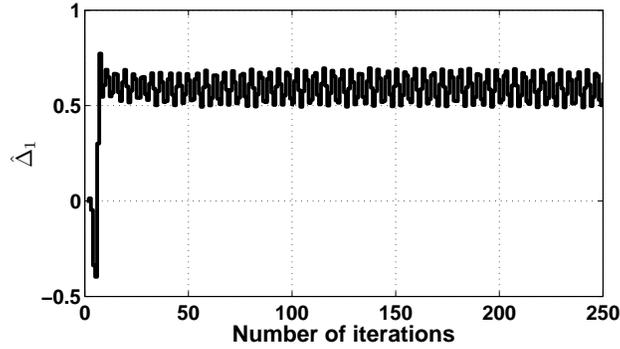}
\label{fig:estimate1_case2}}
  \end{minipage}
  \hfill
  \begin{minipage}{0.5\linewidth}
   \center\subfigure[Estimate of $ \Delta_{2}$ over the learning iterations (MES)]{
\includegraphics[width=1\linewidth]{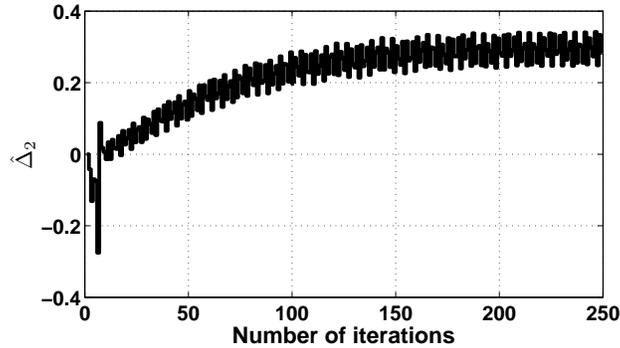}
\label{fig:estimate2_case2}}
  \end{minipage}
  \caption{Cost function and uncertainties estimates- MES algorithm }
  \label{{fig:estimate12_case2}}
\end{figure}

\begin{figure}
   \begin{minipage}{1\linewidth}
\center \hspace{0cm}\subfigure[Obtained vs. desired first angular
trajectory (MES)]{
\includegraphics[width=4.5in,height=2.4in]{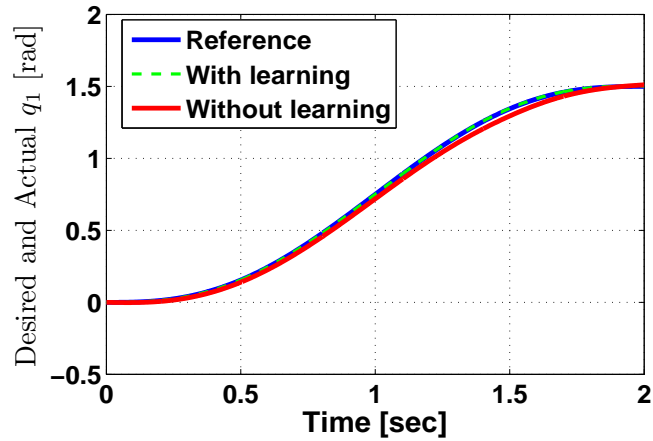}
\label{fig:q1_case2}}
  \end{minipage}
  \hfill
   \begin{minipage}{1\linewidth}
  \center \hspace{0cm}\subfigure[Obtained vs. desired first angular velocity trajectory (MES)]{
\includegraphics[width=4.5in,height=2.4in]{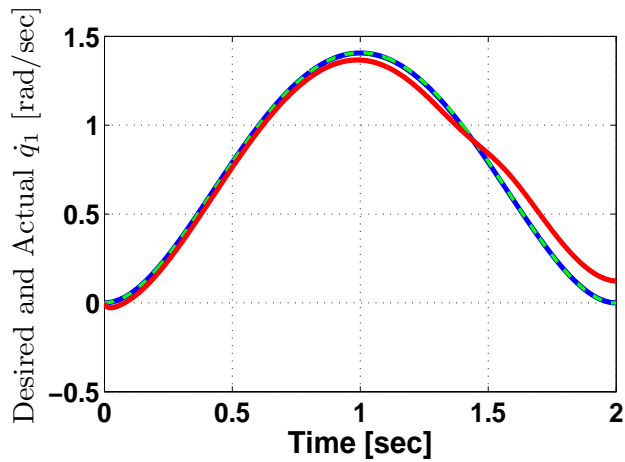}
\label{fig:dotq1_case2}}
  \end{minipage}
  \caption{Obtained vs. desired trajectories (MES) }
  \label{{fig:q1dotq1_case2}}
\end{figure}

The estimates of the two parameters
$\widehat{\Delta}_i$ ($i = 1,2$) are computed using a discrete
version of (\ref{eq:mes}), given by
\begin{equation}
\begin{array}{l}
\displaystyle {x}_i (k+1) ~=~x_i(k) + a_i t_f \sin(\omega_i t_f  k +\frac{\pi}{2}) J(\widehat{\Delta}),\\
\displaystyle  \widehat{\Delta}_i(k+1)~=~x_i(k+1) + a_i
\sin(\omega_i  t_f k -\frac{\pi}{2}) ,\quad i  =1,2
\end{array}
\label{eq:esa}
\end{equation}
where, $k\in\mathbb{N}$ denotes the iteration index,
$x_{i}(0)=\widehat{\Delta}_{i}(0)=0$. We choose the following
learning cost function
\begin{equation}\begin{array}{c} J(\widehat{\Delta})=\int_{0}^{t_f}
(q(\widehat{\Delta})-q_d(t))^TQ_1(q(\widehat{\Delta})-q_d(t))
{\rm{d}} t
\\+\int_{0}^{t_f} (\dot{q}(\widehat{\Delta})-\dot {q}_d(t))^TQ_2(\dot{q}(\widehat{\Delta})-\dot q_d(t))
\rm{d}t,
\end{array}\label{eq:cost}
\end{equation}
where $Q_1>0$ and $Q_2>0$ denote the weight matrices. We implement
the learning parameters:  $a_1=0.1$, $a_2=0.05$,
$\omega_1=7\;rad/sec$, $\omega_2=5\;rad/sec$. The obtained
performance cost function is displayed on Figure
\ref{fig:cost_case2}, where we see that the performance improves
over the learning iterations. The corresponding parameters
estimation profiles are reported on Figures
\ref{fig:estimate1_case2}, and \ref{fig:estimate2_case2}, which
show a quick convergence of the first estimates $\hat{\Delta}_{1}$
to a neighborhood of the actual value. The convergence of the
second estimates $\hat{\Delta}_{2}$ is slower, which is expected
from the ES algorithms when many parameters are estimated at the
same time. One has to underline here, however, that the
convergence speed of the estimates and the excursion around their
final mean values, can be directly fine-tuned by the proper choice
of the learning coefficients $a_{i},\;\omega_{i},\;i=1,2$ in
equation (\ref{eq:esa}). Finally, The tracking performance is
shown on Figures \ref{fig:q1_case2}, \ref{fig:dotq1_case2}, where
we can see that, after learning the actual values of the
uncertainties, the tracking of the desired trajectories is
recovered. We only show the first angular trajectories here,
because the uncertainties affect directly only the acceleration
$\ddot{q}_{1}$, and their effect on the tracking for the second
angular variable is negligible.

\subsection{GP UCB-based uncertainties estimation} \label{sec:constant}
\begin{figure}
 \begin{minipage}{1\linewidth}
   \center\subfigure[Cost function over the learning iterations (GP-UCB)]{
\includegraphics[width=4in,height=2.5in]{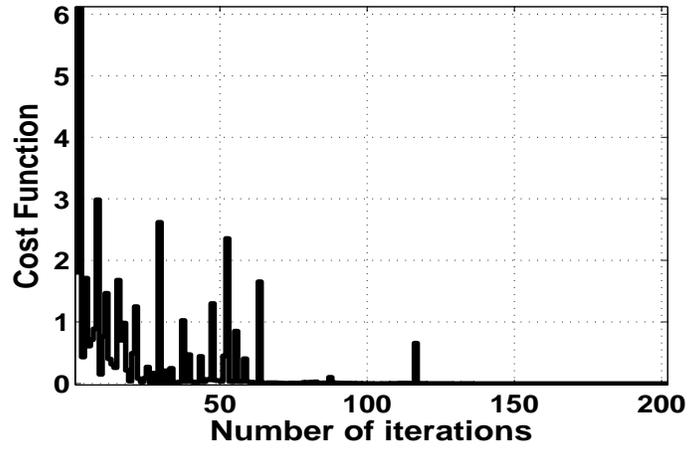}
\label{fig:cost_case1gp}}
  \end{minipage}
  \hfill
  \begin{minipage}{1\linewidth}
   \center\subfigure[Estimate of $ \Delta_{1}$ over the learning iterations (GP-UCB)]{
\includegraphics[width=4in,height=2.5in]{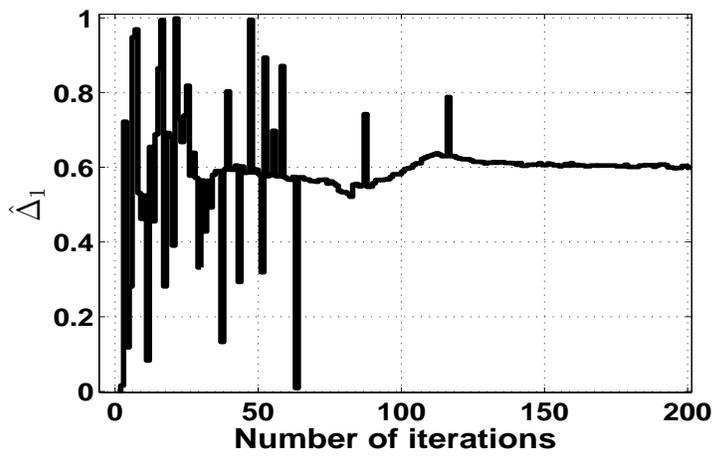}
\label{fig:estimate1_case1gp}}
  \end{minipage}
  \hfill
 \center \begin{minipage}{1\linewidth}
   \center\subfigure[Estimate of $ \Delta_{2}$ over the learning iterations (GP-UCB)]{
\includegraphics[width=4in,height=2.5in]{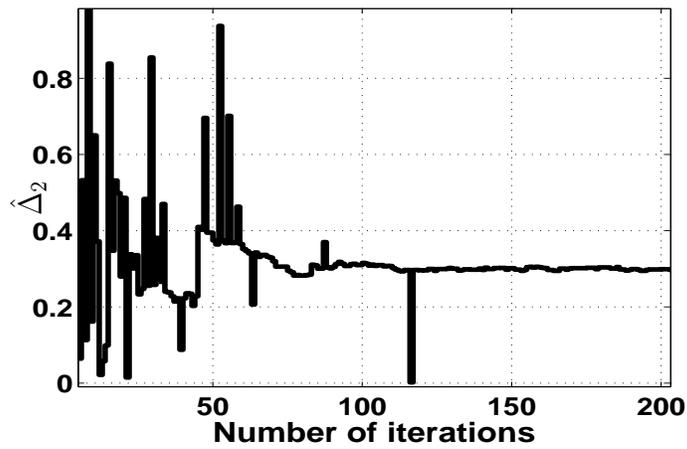}
\label{fig:estimate2_case1gp}}
  \end{minipage}
  \caption{Cost function and uncertainties estimates- (GP-UCB) algorithm }
  \label{{fig:estimate12_case1gp}}
\end{figure}

\begin{figure}
  \begin{minipage}{1\linewidth}
  \center\hspace{-0cm}\subfigure[Obtained vs. desired first angular trajectory (GP-UCB)]{
\includegraphics[width=4.3in,height=2.4in]{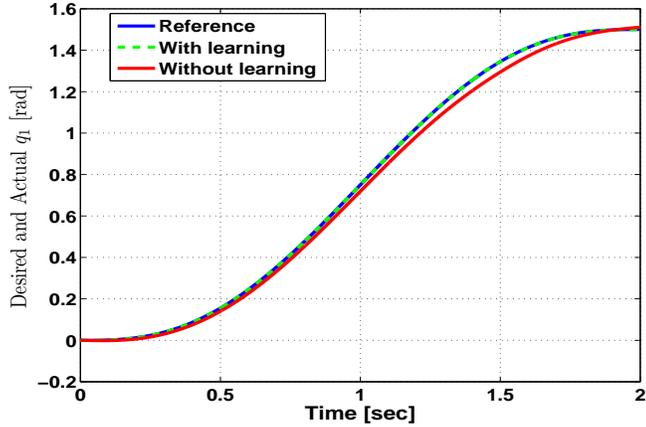}
\label{fig:q1_case1gp}}
  \end{minipage}
  \hfill
  \begin{minipage}{1\linewidth}
  \center \hspace{-0cm}\subfigure[Obtained vs. desired first angular velocity trajectory (GPU-CB)]{
\includegraphics[width=4.3in,height=2.4in]{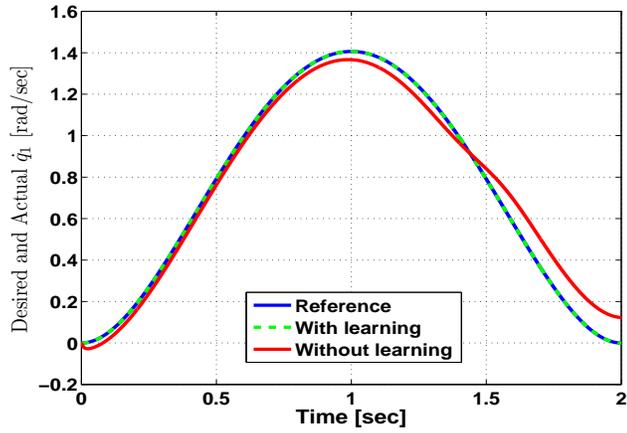}
\label{fig:dotq1_case1gp}}
  \end{minipage}
  \caption{Obtained vs. desired trajectories (GPU-CB)}
  \label{{fig:q1dotq1_case1gp}}
\end{figure}

In this section, to show that the modular ISS-based controller is
independent of the choice of the learning algorithm, we apply the
GP-UCB learning algorithm-based estimator to the same two-links
manipulator example. We apply the algorithm \ref{sec:GP-UCB}, with
the following parameters: $\sigma=0.1$, $l=0.2$, and
$\beta_{t}=2log(\frac{card(D')t^{2}\pi^{2}}{6\delta})$, with
$\delta=0.05$.\\We test the GP-UCB algorithm under the same
conditions as in the previous section. The obtained parameters and
tracking results are reported on figures \ref{fig:cost_case1gp},
\ref{fig:estimate1_case1gp}, \ref{fig:estimate2_case1gp},
\ref{fig:q1_case1gp}, \ref{fig:dotq1_case1gp}. We can see on these
figures that similar to the MES-based adaptive controller, the
uncertainties are well estimated. One could argue that they are
better estimated with the GP-UCB algorithm because there is no
permanent dither signal, which leads to permanent oscillations in
the MES-based learning. The tracking performance is clearly
improved in this case as well, due to the precise estimation of
the parameters.

%
%
%

%
%
\newpage
\section{Conclusion}\label{sec:con}
We have studied the problem of adaptive control for nonlinear
systems which are affine in the control with parametric
uncertainties. For this class of systems, we have proposed the
following controller: We use a modular approach, where we first
design a robust nonlinear controller, designed based on the model
(assuming knowledge of the uncertain parameters), and then
complement this controller with an estimation module to estimate
the actual values of the uncertain parameters. This type of
modular approaches are certainly not new, e.g., the X-swapping
methods. However, the novelty here is that the estimation module
that we propose is based on model-free learning algorithms.
Indeed, we propose to use two learning algorithms, namely, a
multi-parametric extremum seeking algorithm, and a GP-UCB
algorithm, to learn in realtime the uncertainties of the model. We
call the learning approach `model-free' for the simple reason that
it only requires to measure an output signal from the system and
compare it to a desired reference signal (independent of the
model), to learn the best estimates of the model uncertainties. We
have guaranteed the stability (while learning) of the proposed
approach, by ensuring that the model-based robust controller,
leads to an ISS results, which guarantees boundedness of the
states of the closed-loop system, even during the learning phase.
The ISS result together with a convergent learning-algorithm
eventually leads to a bounded output tracking error, which
decreases with the decrease of the estimation error. We believe
that, one of the main advantages of the proposed controller,
comparatively to the existing model-based adaptive controllers, is
that we can learn (estimate) multiple uncertainties at the same
time even if they appear in the model equation in a challenging
way, e.g., linearly dependent uncertainties affecting only one
output, or uncertainties appearing in a nonlinear term of the
model, which are well known limitations of the model-based
estimation approaches. Another advantage of the proposed approach,
is that due to its modular design, one could easily change the
learning algorithm without having to change the model-based part
of the controller. Indeed, as long as the first part of the
controller, i.e., the model-based part, has been designed with a
proper ISS property, one can `plug into it' any convergent
learning model-free algorithm, as demonstrated here by using two
different learning approaches. We reported in this short paper
some preliminary results about using GP-UCB in a modular adaptive
control setting. In a longer journal version of the work, we will
report more detailed comparisons between the MES-based adaptive
controller, the GP UCB-based controller (for example in a more
realistic noisy environment), and some existing model-based
`classical' adaptive controllers, e.g., as found in \cite{IS12}.

\bibliographystyle{IEEEtran}
\footnotesize\bibliography{ACC_16}

\end{document}